\documentclass[floatfix,aps,showpacs,nofootinbib,preprintnumbers,12pt]{revtex4}
\usepackage[T1]{fontenc}
\usepackage[latin1]{inputenc}
\usepackage{float}
\usepackage{graphicx}
\usepackage{epsfig}
\usepackage{latexsym}
\usepackage{amsfonts}
\usepackage{bbold}
\usepackage{graphicx}

\def\Box{\kern1pt\vbox{\hrule height 1.2pt\hbox{\vrule width 1.2pt\hskip 3pt
\vbox{\vskip 6pt}\hskip 3pt\vrule width 0.6pt}\hrule height 0.6pt}\kern1pt}
\def\be{\begin{equation}}
\def\ee{\end{equation}}

\def\be{\begin{equation}}
\def\ee{\end{equation}}
\def\bea{\begin{eqnarray}}
\def\eea{\end{eqnarray}}

\begin{document}

\title{Beyond Spherical Top Hat Collapse}
%

\author{T. S. Pereira$^{1,2}$ R. Rosenfeld$^1$ and A. Sanoja$^1$}

\affiliation{$^1$ Instituto de F\'{\i}sica Te\'orica, Universidade Estadual Paulista\\
Rua Dr. Bento T. Ferraz, 271, 01140-070, S\~ao Paulo, SP, Brazil}
\affiliation{$^2$ Departamento de F\'{\i}sica, Universidade Estadual de Londrina\\
Campus Universit\'ario, 86051-990, Londrina, Paran\'a, Brazil}

\date{\today}


\begin{abstract}
We study the evolution of inhomogeneous spherical perturbations in
the universe in a way that generalizes the spherical top hat collapse
in a straightforward manner. For that purpose we will derive a dynamical
equation for the evolution of the density contrast in the context
of a Lemaître-Tolman-Bondi metric and construct solutions with and
without a cosmological constant for the evolution of a spherical perturbation
with a given initial radial profile.
\end{abstract}

\maketitle

\section{Introduction}

The formation of large scale structure in the universe is one of the
most promising probes that will be used to determine fundamental cosmological
parameters in future extra-galactic surveys, such as the imminent
Dark Energy Survey (DES) \cite{Abbott:2005bi}, the Large Synoptic
Survey Telescope (LSST) \cite{Tyson:2003kb} and the Euclid survey
\cite{Euclid:2008eu}.

The basic picture for large scale structure formation is that small
perturbations generated by quantum fluctuations in the inflationary
epoch grow in the dark matter to form gravitational potential wells
where the baryons later fall into. The linear part of this process
for dark matter is well understood and described by linear perturbation
theory \cite{Bardeen:1980kt,Kodama:1985bj}. However, the nonlinear
stages are not amenable to these perturbative methods even in the
case when only dark matter is present. One then usually resorts to
large numerical N-body simulations in order to obtain, \textit{e.g.}
the mass distribution of large dark matter haloes. Unfortunately,
these simulations are very costly and it is desirable that a simple,
approximate semi-analytical approach could be used to estimate the
properties of dark matter haloes.

One such model is the so-called spherical collapse model \cite{Gunn:1972sv}.
In this extremely simple model, a spherically symmetric region of
homogeneous overdensity, called {}``top-hat'' density profile, evolves
inside a homogeneous expanding Universe. Symmetry arguments show
that one can regard the overdense region as a mini-universe of positive
curvature. Hence the 
Raychaudhury and continuity equations can be used to
evolve the density and radius of the spherical region \cite{Padmanabhan,Fosalba:1997tn}.
When combined with the Press-Schechter theory \cite{Press:1973iz}
this framework provides a statistical basis for structure formation
from which the number density of dark matter haloes can be estimated.

The spherical collapse model has been recently used to study structure
formation in models with a simple Yukawa-type modification of gravity
\cite{Martino:2008ae}, in the so-called $f(R)$ models \cite{Schmidt:2008tn},
in braneworld cosmologies \cite{Schmidt:2009yj}, in models which
allow for dark energy fluctuations \cite{Mota:2004pa,Nunes:2004wn,Nunes:2005fn,Horellou:2005qc,Manera:2005ct,Schaefer:2007nf,Abramo:2007iu,Abramo:2008ip,Basilakos:2009mz,Pace:2010sn,Wintergerst:2010ui}
and in the so-called chameleon models \cite{Brax:2010tj}. 
The purpose of this work is to relax the assumption of a top-hat
profile and study the nonlinear evolution of an inhomogeneous spherical
perturbation in a fully relativistic framework.

\section{Evolution of an inhomogeneous spherical perturbation}

In the case of an inhomogeneous perturbation one can not use a Friedmann-Robertson-Walker
(FRW) metric which is valid only for a homogeneous matter distribution.
The most general spherically symmetric metric is given by: \begin{equation}
ds^{2}=e^{A(r,t)}dt^{2}-e^{B(r,t)}dr^{2}-R^{2}(r,t)d\Omega^{2}\end{equation}
 where $t$ is the cosmic time, $r$ is the comoving radial coordinate
and $d\Omega$ is the solid angle element. The metric is determined
by three functions: $A(r,t)$, $B(r,t)$ and $R(r,t)$, where this
last function is known as the areal radius since the area of a surface
with a given time and comoving radial coordinates is given by $S=4\pi R(r,t)^{2}$.
For a nice discussion on spherically symmetric inhomogeneous models
see, {\it e.g.}, chapter 18 in \cite{Plebanski:2006sd}.

We assume a single perfect fluid with energy-momentum tensor given
by \begin{equation}
T_{\;\; b}^{a}=(\rho+p)u^{a}u_{b}-p\delta_{\;\; b}^{a}\end{equation}
 where $\rho(r,t)$ and $p(r,t)$ are the energy density and pressure
of the fluid and $u^{a}$ is its $4-$velocity, with $u^{a}u_{a}=1$.
In a comoving reference frame $u=(e^{-A/2},\vec{0})$.

In this metric, Einstein equations with a cosmological constant, $G_{ab}-\Lambda g_{ab}=8\pi GT_{ab}$,
can be written as:

\begin{eqnarray}
e^{-A}\left(\frac{\dot{R}^{2}}{R^{2}}+\frac{\dot{B}\dot{R}}{R}\right)-e^{-B}\left(2\frac{R''}{R}+\frac{R'^{2}}{R^{2}}-\frac{B'R'}{R}\right)+\frac{1}{R^{2}} & = & 8\pi G\rho+\Lambda\label{eq:ee1}\end{eqnarray}
 \begin{equation}
e^{-B}\left(2\frac{\dot{R}'}{R}-\frac{\dot{B}R'}{R}-\frac{A'\dot{R}}{R}\right)=0\label{eq:ee2}\end{equation}
 \begin{equation}
e^{-A}\left(2\frac{\ddot{R}}{R}+\frac{\dot{R}^{2}}{R^{2}}-\frac{\dot{A}\dot{R}}{R}\right)-e^{-B}\left(\frac{R'^{2}}{R^{2}}+\frac{A'R'}{R}\right)+\frac{1}{R^{2}}=-8\pi Gp+\Lambda\label{eq:ee3}\end{equation}
 \begin{eqnarray}
\frac{e^{-A}}{4}\left(4\frac{\ddot{R}}{R}-2\frac{\dot{A}\dot{R}}{R}+2\frac{\dot{B}\dot{R}}{R}+2\ddot{B}+\dot{B}^{2}-\dot{A}\dot{B}\right)-\nonumber \\
-\frac{e^{-B}}{4}\left(4\frac{R''}{R}+2\frac{A'R'}{R}-2\frac{B'R'}{R}+2A''+A'^{2}-A'B'\right) & = & -8\pi Gp+\Lambda\label{eq:ee4}\end{eqnarray}
 where dots and primes refer to partial derivatives with respect to
time and space, respectively.

Conservation of the energy-momentum tensor, $\nabla_{a}T_{\,\, b}^{a}=0$,
in this spacetime results in the following equations: \begin{equation}
\dot{B}+4\frac{\dot{R}}{R}+2\frac{\dot{\rho}}{\rho+p}=0\label{eq:c1}\end{equation}
 \begin{equation}
A'+2\frac{p'}{\rho+p}=0\,.\label{eq:ec2}\end{equation}

Equations (\ref{eq:ee1}) and (\ref{eq:ee3}), combined with Eq. (\ref{eq:ee2}),
can be written as two conservation equations for the so-called \textit{active}
mass $m(r,t)$: \begin{eqnarray}
\frac{2m'(r,t)}{R^{2}R'} & = & 8\pi G\rho\\
\frac{2\dot{m}(r,t)}{R^{2}\dot{R}} & = & -8\pi Gp\end{eqnarray}
 with \begin{equation}
2m(r,t)=e^{-A}R\dot{R}^{2}-e^{-B}RR'^{2}+R-\frac{1}{3}\Lambda R^{3}\,.\end{equation}

We will focus on pressureless dark matter, in which case $p=0$ and
\begin{equation}
\dot{m}(r,t)=0\quad\Rightarrow\quad m=m(r).\end{equation}
In this case the metric takes the form of the Lema\^{\i}tre-Tolman-Bondi (LTB)
metric \cite{Lemaitre:1933gd,Tolman:1934za,Bondi:1947av}: 
\begin{equation}
ds^{2}=dt^{2}-\frac{R'^{2}}{1+f(r)}dr^{2}-R^{2}(r,t)d\Omega^{2}
\end{equation}
where the curvature $f(r)$ is determined by $m(r)$ and $R$ 
\begin{equation}
f(r)=\dot{R}^{2}-\frac{2m(r)}{R}-\frac{1}{3}\Lambda R^{3}.
\end{equation}

Models with large inhomogeneities described by a LTB metric, such
as the so-called {}``Hubble bubble'' model, have been used recently
as an alternative explanation to the apparent acceleration of the
universe (see, \textit{e.g} \cite{Kolb:2005da,Enqvist:2007vb,GarciaBellido:2008nz,Paranjape:2008ai}).
Our goal here is to study the cosmological evolution of such an inhomogeneity.

The dynamical equations for the LTB model can be written as a generalization
of the Friedmann equation \begin{equation}
\frac{\dot{R}^{2}}{R^{2}}=\frac{2m(r)}{R^{3}}+\frac{f(r)}{R^{2}}+\frac{\Lambda}{3}\end{equation}
 and a generalization of the {}``acceleration'' equation \begin{equation}
\frac{\ddot{R}}{R}=-\frac{m(r)}{R^{3}},\end{equation}
 which is independent of the curvature, just like the FRW case. These
equations reduce to the usual FRW equations if we set \begin{equation}
R(r,t)=a(t)r\end{equation}
 and \begin{equation}
f(r)=-kr^{2},\end{equation}
 where $k$ is the spatial curvature.

To describe the evolution of the inhomogeneous density perturbation
we define the density contrast
\begin{equation}
\delta\left(r,t\right)\equiv\frac{\rho\left(r,t\right)-\overline{\rho}\left(t\right)}{\overline{\rho}\left(t\right)}\label{eq:definicion1}\end{equation}
 where $\rho(r,t)$ is the energy density inside the perturbation
and $\overline{\rho}(t)$ is the energy density of the 
unperturbed expanding background.

The dynamical equation for $\delta(r,t)$ can be deduced from Eq.
(\ref{eq:c1}), which for a pressureless fluid becomes
\begin{equation}
\dot{\rho}+3h\rho=0\label{eq:continuidad}\end{equation}
 where $h\equiv\dot{b}/b$ and $b\equiv(R^{2}R')^{1/2}$, together
with the expanding background continuity equation:
\begin{equation}
\dot{\overline{\rho}}+3H\overline{\rho}=0\,\label{eq:continuidad2}\end{equation}
 where $H=\dot{a}/a$. 
Taking the derivative of Eq. (\ref{eq:definicion1})
with respect to time we obtain
\begin{equation}
\dot{\rho}=\dot{\overline{\rho}}\left(1+\delta\right)+\overline{\rho}\dot{\delta}\,,\label{eq:Derivada1}\end{equation}
 which can be rewritten as
\begin{equation}
\dot{\delta}=3\left(H-h\right)\left(1+\delta\right)\label{eq:Derivada2}\end{equation}
 where we have used Eqs. (\ref{eq:continuidad}) and (\ref{eq:continuidad2}).

Deriving Eq. (\ref{eq:Derivada2}) with respect to time we obtain
\begin{equation}
\ddot{\delta}=3\left(\dot{H}-\dot{h}\right)\left(1+\delta\right)+3\left(H-h\right)\dot{\delta}\,.\label{eq:Derivada3}\end{equation}
If we now combine Eqs. (\ref{eq:ee1})-(\ref{eq:ee4}) with the equations
\begin{equation}
\dot{H}=-\frac{3}{2}H^{2}+\frac{\Lambda}{2}\label{eq:hubble}\end{equation}
 and \begin{equation}
\dot{h}=4\pi G\rho+\frac{2}{3R}\left(\frac{f}{R}+\frac{f'}{R'}\right)+\Lambda-3h^{2}\label{eq:hubble1}\end{equation}
 Eq. (\ref{eq:Derivada3}) becomes the nonlinear differential equation
for the evolution of the density perturbation:
\begin{equation}
\ddot{\delta}+2H\dot{\delta}-4\pi G\bar{\rho}\delta\left(1+\delta\right)-\frac{4}{3}\frac{\dot{\delta}^{2}}{1+\delta}=\frac{\Lambda}{2}\left(1+\delta\right)+\frac{2}{3}\left[\frac{\partial}{\partial t}\ln\frac{R}{R'}\right]^{2}\left(1+\delta\right)\,.\label{eq:evolutiondensity}\end{equation}

Note that for a homogeneous density perturbation with a top-hat profile
the areal radius is $R=ra$. In this case, the last term on the right
hand side of Eq. (\ref{eq:evolutiondensity}) vanishes, reducing this
equation to the well known evolution equation (see, \textit{e.g.} \cite{Abramo:2007iu})
in the case of a single pressureless fluid.

Eq. (\ref{eq:evolutiondensity}) generalizes 
the evolution of $\delta$ for spherical perturbations
with arbitrary initial radial profiles. Note, however, that for the
simple case of pressureless dark matter, the solution of this equation
can be obtained directly from Eqs. (\ref{eq:continuidad}) and (9),
which is what we will do in the following.

Consider initially a radial density profile expanding with the background
\begin{equation}
R_{i}=ra_{i}\label{eq:condicion1}\end{equation}
 which is characterized by a time-independent function $g(r)$ \begin{equation}
\rho_{i}\left(r\right)=\bar{\rho_{i}}\left\{ 1+\delta_{i}\left(r\right)\right\} =\bar{\rho_{i}}g\left(r\right)\,.\label{eq:condicion2}\end{equation}

Since Eq. (9) in LTB is time-independent, we can set
\begin{equation}
4\pi GR_{i}^{2}R_{i}'\rho_{i}\left(r\right)=4\pi GR^{2}R'\rho\,.\label{eq:solution}\end{equation}
 Furthermore, since $\bar{\rho}_{0}=\bar{\rho}a^{3}$ and $\rho=\bar{\rho}\left\{ 1+\delta\right\} $
we obtain
\begin{equation}
\delta\left(r,t\right)=a\left(t\right)^{3}\frac{g\left(r\right)r^{2}}{R^{2}R'}-1\,.\label{eq:solution1}\end{equation}

Defining a dimensionless areal radius \begin{equation}
\widetilde{R}=\frac{R}{r}\label{eq:definition2},\end{equation}
 Eq. (\ref{eq:solution1}) becomes \begin{equation}
\delta\left(r,t\right)=a\left(t\right)^{3}\frac{g\left(r\right)}{\widetilde{R}^{2}\left[\widetilde{R}+r\widetilde{R}'\right]}-1,\end{equation}
which is, of course, the formal solution to the nonlinear evolution
equation Eq. (\ref{eq:evolutiondensity}).

Before we find the evolution of the density contrast we need to solve
the dynamical equations of the background FRW metric for $a(t)$ and
of the LTB metric for $\widetilde{R}(r,t)$.

\section{Solutions of the LTB model for a given initial density profile}

We consider initially solutions of the LTB model applied to an inhomogeneous
density perturbation with an initial density profile embedded in a
Einstein-de Sitter and $\Lambda$CDM background.

\subsection{Einstein-de Sitter background}

Our first goal is to study the evolution of an initial spherically
symmetric perturbation in a dark matter dominated universe with an
Einstein- de Sitter (EdS) background in the context of General Relativity.

We assume that at an arbitrary initial time $t=t_{i}$, the whole
universe follows the background expansion with $R(r,t_{i})=a(t_{i})r$,
but has an initial small density perturbation specified by a profile
$g(r)$: \begin{equation}
\rho(r,t_{i})=\bar{\rho}(t_{i})g(r)\end{equation}
 where $\bar{\rho}(t_{i})$ is the initial homogeneous background
density.

Since $m(r)$ is time-independent it can be computed as: \begin{equation}
m(r)=4\pi Ga_{i}^{3}\bar{\rho}_{i}\int_{0}^{r}dr'r'^{2}g(r').\end{equation}
 Defining \begin{equation}
h(r)=\frac{3}{r^{3}}\int_{0}^{r}dr'r'^{2}g(r')\end{equation}
 we can write \begin{equation}
m(r)=\frac{4\pi G}{3}\bar{\rho}_{0}r^{3}h(r),\end{equation}
 where the background density today is $\bar{\rho}_{0}=a_{i}^{3}\bar{\rho}_{i}$.

Using time in units of the Hubble age today \begin{equation}
\tilde{t}=tH_{0}\end{equation}
 and distance in units of the comoving radius \begin{equation}
\tilde{R}=\frac{R}{r}\end{equation}
 we will solve the acceleration equation \begin{equation}
\frac{\partial^{2}\tilde{R}}{\partial\tilde{t}^{2}}=-\frac{1}{2}\Omega_{0}\tilde{R}^{-2}h(r)\end{equation}
 for a given value of $r$ and initial conditions \begin{equation}
\left(\frac{\partial\tilde{R}}{\partial\tilde{t}}\right)_{i}=\sqrt{a_{i}^{-1}\Omega_{0}+(1-\Omega_{0})}\end{equation}
 and $\tilde{R}_{i}=a_{i}$.

Hence, given the initial conditions, the evolution of the inhomogeneous
perturbation is completely determined by its initial profile. As an
example of our method, we choose a Gaussian profile with width given
by $r_{c}$ with a sharp cutoff at $r=r_{c}$ characterizing the comoving
size of the perturbation: \begin{equation}
g(r)=1+\delta_{i}e^{-r^{2}/r_{c}^{2}}\Theta(r_{c}-r)\end{equation}
 where $\Theta(x)$ is the usual Heaviside function. We have started
our evolution from $a_{i}=10^{-5}$ and chose $\delta_{i}=10^{-4.552}$,
which results in the collapse today ($a=1$ when $\tilde{t}=0.667$)
of the center of the perturbation. Exact solutions for the dynamical
LTB equations in this case exist in a parametric form \cite{Plebanski:2006sd}
and we used them to check our numerical method.

\begin{figure}[h,t,b]
\begin{centering}
\includegraphics[clip,scale=0.8]{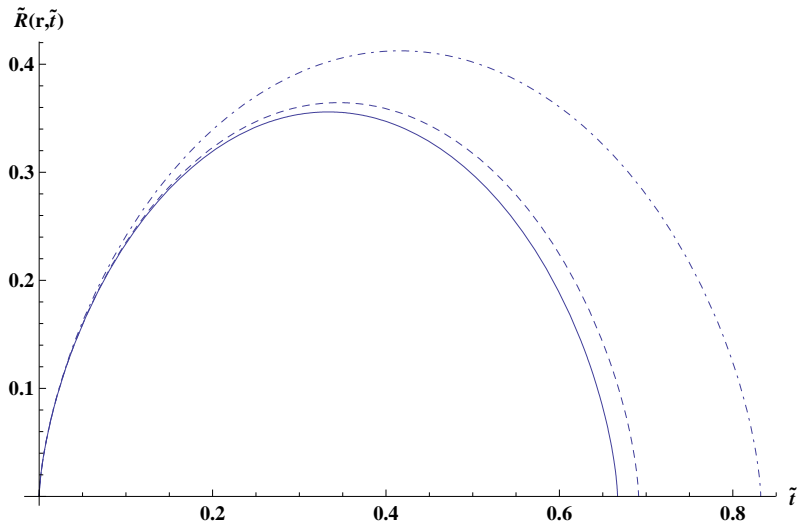}
\caption{\label{Rtilde} \small \sf
Areal radius $\tilde{R}(r,\tilde{t})$
as a function of time $\tilde{t}$ for $r/r_{c}=0$ (solid line),
$0.2$ (dashed line) and $0.5$ (dot-dashed line).
}
\par\end{centering}
\end{figure}

In Fig.(\ref{Rtilde}) we show the time evolution of $\tilde{R}(r,t)$
for different values of the comoving radius $r$ in units of $r_{c}$.
We can see that different shells collapse at different times, with
the outer shells collapsing later. Hence, there is no shell crossing
in this case, as expected since the density profile is decreasing
radially.

%


Interestingly, the presence of a density perturbation in our example
of an EdS universe, even when initially localized, makes the whole
universe to eventually collapse. For instance, a region with a comoving
radius of $r=2r_{c}$ collapses at $\tilde{t}=34.1$. Therefore, the
expansion rate of the universe at this radius is slightly different
from the background EdS.

\begin{figure}[h]
\begin{centering}
\includegraphics[clip,scale=1.0]{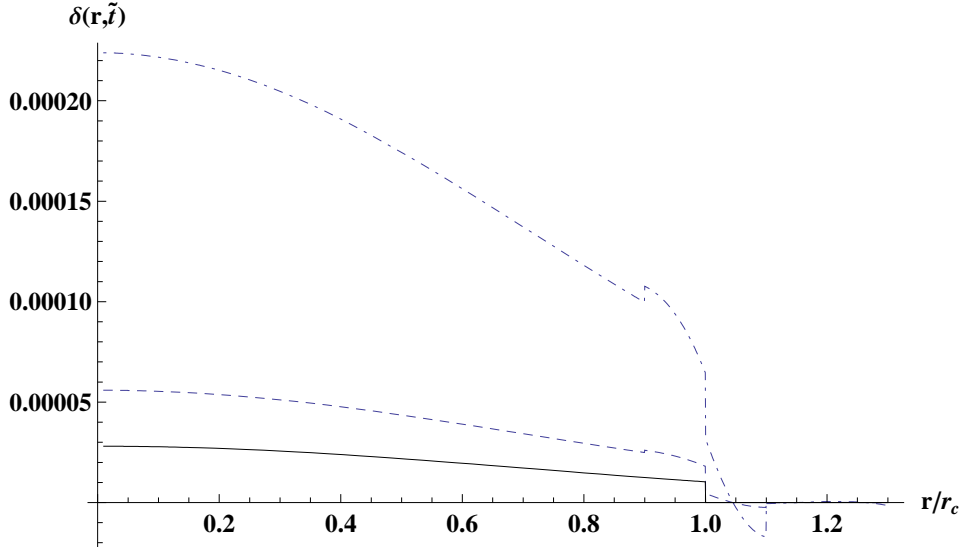}
\caption{\label{DeltaInitial} \small \sf Density contrast $\delta(r,\tilde{t})$
as a function of comoving radius in units of $r_{c}$ for the initial
contrast (solid line), density contrast at $\tilde{t}=10^{-7}$ (dashed
line) and $\tilde{t}=10^{-6}$ (dot-dashed line).}
\par\end{centering}
\end{figure}

In Fig.(\ref{DeltaInitial}) we show the initial instants of the evolution
of the profile of the density contrast $\delta(r,\tilde{t})$. The
behavior of the perturbation near the sharp boundary is a consequence
of numerical instabilities that do not appear in the exact solution.

%

\begin{figure}[h]
\begin{centering}
\includegraphics[clip,scale=1.0]{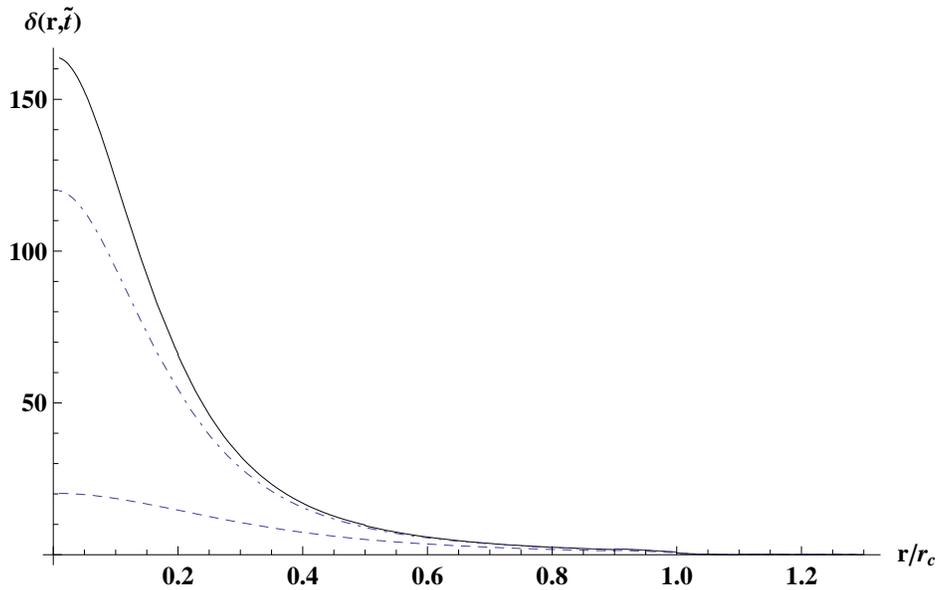}
\caption{\label{DeltaFinal}\sf \small Density contrast $\delta(r,\tilde{t})$
as a function of comoving radius in units of $r_{c}$ for $\tilde{t}=0.5$
(dashed line), density contrast at $\tilde{t}=0.60$ (dot-dashed line)
and $\tilde{t}=0.61$ (solid line).}
\par\end{centering}
\end{figure}

When one is still in the linear regime, the shape of the profile does
not change significantly. For times close to the collapse of the center,
the density profiles approach Gaussian profiles with different widths,
as shown in Fig.(\ref{DeltaFinal}).

%


\subsection{$\Lambda$CDM background}

We now proceed to study the evolution of an inhomogeneous spherical
density perturbation in a background with dark matter and cosmological
constant, where $\overline{\rho}\left(t\right)=\frac{\overline{\rho}_{m}^{0}}{a^{3}}+\rho_{\Lambda}$
and $\Omega_{m}^{0}+\Omega_{\Lambda}^{0}=1$.

As the cosmological constant is not perturbed, it only affects the
behavior of the density contrast through changing the evolution of
$a\left(t\right)$ and $\widetilde{R}\left(r,t\right)$. We solve
the LTB equations with a cosmological constant, assuming that at an
arbitrary initial time $t_{i}=0$, the whole universe follows the
background expansion with $R(r,t_{i})=a(t_{i})r$ but it has an initial
small density perturbation specified by the profile $g(r)$.

The function mass $m(r)$ now becomes \begin{equation}
m\left(r\right)=\frac{1}{2}H_{0}^{2}\Omega_{m}^{0}r^{3}h\left(r\right)\,.\label{eq:mass}\end{equation}

Using time in units of the Hubble age today and distance in units
of the comoving radius, we will solve the acceleration equation \begin{equation}
\frac{\partial^{2}\widetilde{R}}{\partial\widetilde{t}^{2}}=-\frac{1}{2}\Omega_{m}^{0}\widetilde{R}^{-2}h\left(r\right)+\Omega_{\Lambda}^{0}\widetilde{R}\label{eq:equationcc}\end{equation}
 for a given value of $r$ and initial conditions \begin{equation}
\left(\frac{\partial\widetilde{R}}{\partial\widetilde{t}}\right)_{i}=\sqrt{\frac{\Omega_{m}^{0}}{a_{i}}+\Omega_{\Lambda}^{0}a_{i}^{2}}\label{eq:equationic}\end{equation}
 and $\widetilde{R}_{i}=a_{i}$. In our example below we will use
$\Omega_{m}^{0}=0.23$ and $\Omega_{\Lambda}^{0}=0.77$, in which
case $\tilde{t}=1.037$ today.

We should point out that also in this case there are exact solutions
in terms of implicit Weierstrass-p functions for which the implementation
has to be carried semi-analytically \cite{1965PNAS...53....1O}. Since
this implementation is perhaps as intricate as solving the equations
numerically, we have opted for the latter procedure, for simplicity.

\begin{figure}[h]
\begin{centering}
\includegraphics[clip,scale=0.8]{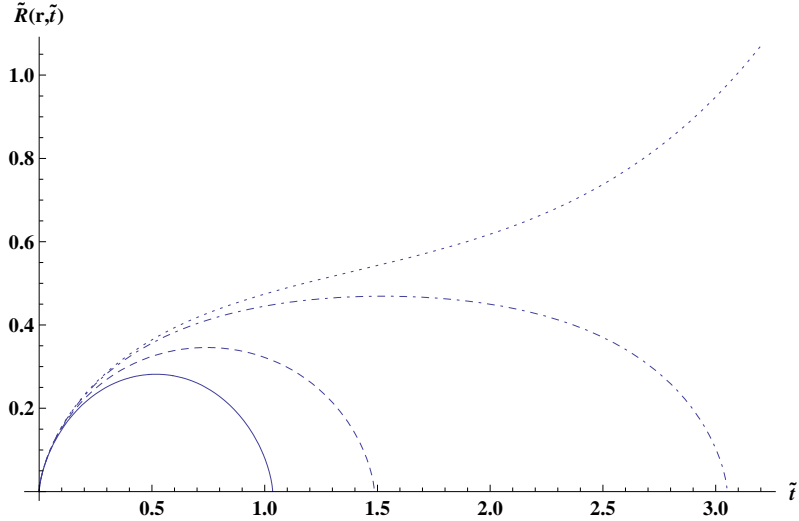}
\caption{\label{RtildeLCDM} \sf \small Areal radius $\tilde{R}(r,\tilde{t})$
as a function of time $\tilde{t}$ for $r/r_{c}=0.01$ (solid line),
$0.5$ (dashed line), $0.7$ (dot-dashed line) and $0.75$ (dotted
line).}
\par\end{centering}
\end{figure}

In Fig. (\ref{RtildeLCDM}) we show that the time evolution of $\tilde{R}(r,t)$
has the same qualitative behavior as in the EdS case, except that
from a certain radius on (around $r=0.7r_{c}$ in our example) the
effect of the cosmological constant halts the collapse. This is in
agreement with the recent analysis of \cite{Mimoso:2009wj}, where
it was found a dividing shell separating expanding and collapsing
regions in general models.

%

\begin{figure}[h]
\begin{centering}
\includegraphics[clip,scale=1.0]{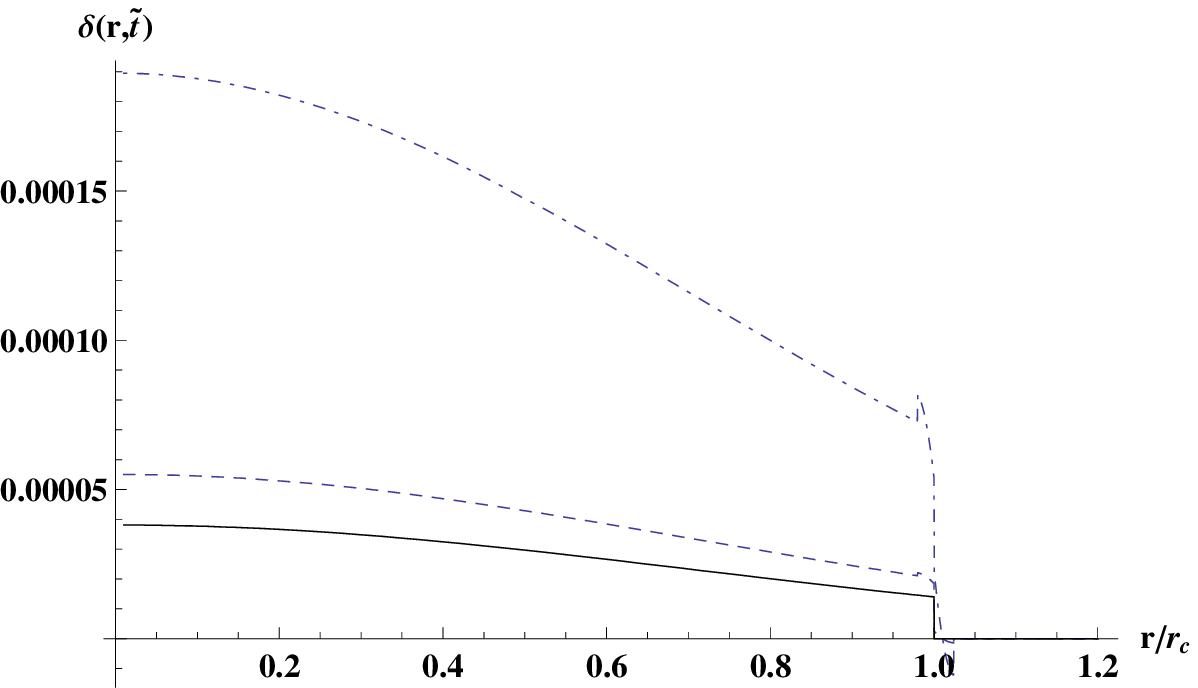}
\par\end{centering}
\caption{\label{DeltaInitialLCDM} \textsf{\small Density contrast $\delta(r,\tilde{t})$
as a function of comoving radius in units of $r_{c}$ for the initial
contrast (solid line), density contrast at $\tilde{t}=10^{-7}$ (dashed
line) and $\tilde{t}=10^{-6}$ (dot-dashed line).}}
\end{figure}
\begin{figure}[h]
\begin{centering}
\includegraphics[clip,scale=1.1]{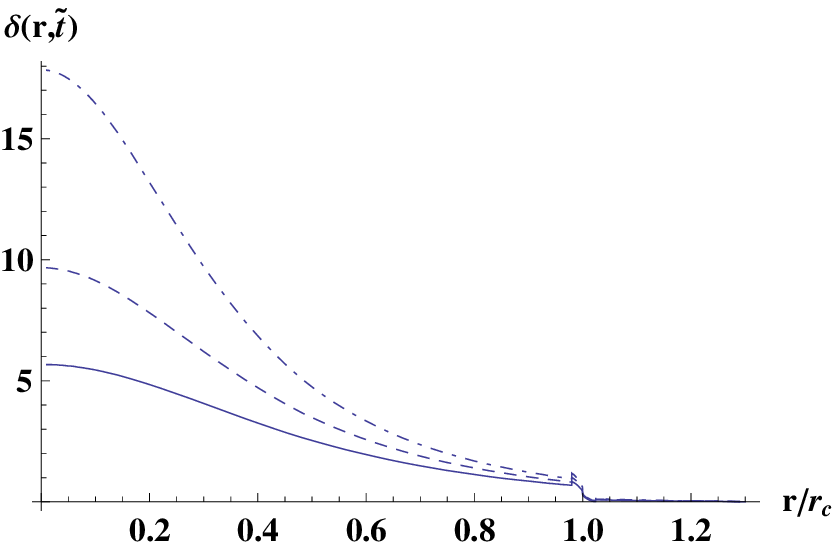}
\par\end{centering}
\caption{\label{DeltaFinalLCDM} \textsf{\small Density contrast $\delta(r,\tilde{t})$
as a function of comoving radius in units of $r_{c}$ at $\tilde{t}=0.5$
(solid line), $\tilde{t}=0.6$ (dashed line) and $\tilde{t}=0.7$
(dot-dashed line).}}
\end{figure}

The initial evolution of perturbations in matter is shown in Fig.(\ref{DeltaInitialLCDM}).
Again we show the initial instants of the evolution of the profile
of the density contrast $\delta(r,\tilde{t})$. The small bump around
$r=r_{c}$ is due to numerical instabilities and appears because of
the sharp boundary in the density profile. Notice that the perturbation
profile rapidly evolves to a Gaussian in the nonlinear regime, as
can be seen in Fig.(\ref{DeltaFinalLCDM}).

%


%


\section{Conclusion}

In this work we have implemented a simple generalization of the
spherical top-hat collapse model including a general initial profile
for the perturbation. We compute the nonlinear evolution of the density
perturbation. We chose as an example a simple Gaussian profile with
a sharp cutoff that separates the perturbed region from the background.
We showed that the density perturbations evolve differently in an
EdS and a $\Lambda$CDM background, where in the latter case there
is a dividing shell between an expansion and a contraction region
inside the perturbation. 
One should notice that from the usual definition of the quantity $\delta_c(z)$, which is
the linearly evolved density contrast with an initial condition such that the nonlinear
collapse occurs at a redshift $z$, that its value in principle depends only on the
background cosmology, being independent of the density profile.

We plan to extend our method to more realistic
models including interacting fluids with pressure, where no closed form solutions can
be obtained.

\section*{Acknowledgments}

We are grateful to Ronaldo Batista and David Mota for useful comments.
We also thank Nelson Nunes for pointing out reference \cite{Mimoso:2009wj}
to us. The work of AS is supported by a CAPES doctoral fellowship.
TSP thanks FAPESP for financial support. The work of RR is partially supported by
a FAPESP project and a CNPq fellowship.

\section*{References}

\bibliographystyle{h-physrev3}
\bibliography{LTBbiblio}


\end{document}